\documentclass[fleqn,twoside]{article}
\usepackage{espcrc2}

\usepackage{graphicx}
\usepackage[figuresright]{rotating}

\newcommand{\AmS}{{\protect\the\textfont2
  A\kern-.1667em\lower.5ex\hbox{M}\kern-.125emS}}

\title{Charmonia above the Deconfinement
Phase Transition\thanks{Presented by M. Asakawa}}

\author{M. Asakawa\address{Department of Physics, 
        Kyoto University, Kyoto 606-8502, Japan} and
        T. Hatsuda\address{Department of Physics,
        University of Tokyo, Tokyo 113-0033, Japan} 
        }
       
\begin{document}

\begin{abstract}

Analyzing correlation functions of charmonia
at finite temperature ($T$)  
on $32^3\times$(32$-$96) anisotropic lattices 
by the maximum entropy method (MEM),
we find that $J/\psi$ and $\eta_c$ survive
as distinct resonances in the plasma
even up to $T \simeq 1.6 T_c$ and that they eventually dissociate 
between $1.6 T_c$ and $1.9 T_c$ ($T_c$ is the 
critical temperature of deconfinement).
This suggests that the deconfined plasma is non-perturbative
enough to hold heavy-quark bound states.
The importance of having sufficient number of
temporal data points in the MEM analysis is also emphasized.

\vspace{1pc}
\end{abstract}

\maketitle

\setcounter{footnote}{0}

\section{INTRODUCTION}

Whether hadrons survive even in the deconfined quark-gluon plasma
is one of the key questions in   
quantum chromodynamics (QCD). 
This problem was first examined 
in \cite{HK85} and \cite{DT85} in different contexts.
The fate of the heavy mesons such as $J/\psi$ 
in the deconfined plasma  was also investigated in  
a phenomenological potential picture taking into account the Debye 
screening \cite{MS86}.
In general, there is no a priori reason to believe that
dissociation of the bound states should take place exactly  
at the phase transition point \cite{NSR85}.

From the theoretical point of view, the 
spectral function (SPF) at finite temperature $T$, which
has all the information of in-medium 
hadron properties, is a key quantity to be studied. 
Recently, the present authors have shown \cite{nah99} that 
the first-principle lattice QCD simulation
of SPFs is possible  by utilizing the maximum
entropy method (MEM). 
We have also formulated the basic concepts and applications of MEM
on the lattice  at $T=0$ and $T\neq 0$ in \cite{ahn01}.
To draw the conclusion with a firm ground,
in the following we put special emphases on
[I] the MEM error analysis of the resultant SPFs
and [II] the sensitivity of the SPFs to $N_{\rm data}$
(the number of temporal data points adopted in MEM).
These tests are crucial to prevent fake generation
and/or smearing of the peaks and must be always carried out
as emphasized in \cite{ahn01,ahn02,ah03}. 

\section{LATTICE SIMULATION}

The applications of MEM to $T\neq 0$ system have been known to be
a big challenge \cite{ahn01}.
The difficulty originates from the fact that
the temporal lattice size $L_{\tau}$ is restricted as
$L_{\tau}=1/T=N_{\tau} a_{\tau}$,
where $a_\tau$ ($N_{\tau}$) is the
temporal lattice spacing (number of temporal lattice sites).
Because of this, it becomes more difficult to keep enough $N_{\rm data}$
to obtain reliable SPFs as $T$ increases.
Thus, simulations up to a few times $T_c$ with
$N_{\rm data} $ as large as 30 \cite{ahn02}
inevitably require an anisotropic lattice.
 
On the basis of the above observation, we have
carried out  quenched simulations  with $\beta=7.0$
on $32^3\times N_\tau $ anisotropic lattice with
$N_\tau = 32,~40,~46,~54$, and $96$.
The renormalized anisotropy is
$\xi = a_{\sigma}/a_{\tau}=4.0$ with $a_{\sigma}$ being the
spatial lattice spacing.
We take the naive plaquette gauge action and the standard
Wilson quark action. Corresponding bare anisotropy $\xi_0=3.5$ is
determined from the data given in \cite{klassen,karsch_aniso}.
The fermion anisotropy
$\gamma_F \equiv \kappa_\tau /\kappa_\sigma = 3.476$ with
$\kappa_\sigma = 0.08285$ ($\kappa_\tau = 0.2880$) being the
the spatial (temporal) hopping parameter
is determined by comparing the temporal and
spatial effective masses of
pseudoscalar and vector mesons on a
$32^2 \times 48\times 128$ lattice. 
$a_\tau = a_{\sigma}/4 = 9.75\times 10^{-3}$ fm
is determined  from the $\rho$ meson mass in the chiral limit.
The masses determined on $T=0$ ($32^2 \times 48\times 128$)
lattice are
$m_{J/\psi}=3.10$ GeV and $m_{\eta_c}=3.03$ GeV.
If the temporal distance between the source and the
sink is closer than $\xi a_{\tau}$,
lattice artifact due to anisotropy  would appear
in the SPFs for $\omega \ge \pi/\xi a_\tau$.
To avoid this, we exclude
the six points near the edge ($\tau_i = 1, 2, 3$ and
$N_\tau- 3, N_\tau -2, N_\tau-1$) and adopt
the points $\tau_i= 4, 5, \cdots$ and $N_{\tau}-4, N_{\tau}-5, \cdots$
until we reach the total number of points
$N_{\rm data} (<N_{\tau}-7)$\footnote{For further details
of the lattice simulation, see \cite{ah03}.}.

\section{RESULTS}

\begin{figure}[bht]
\begin{flushleft}
\includegraphics[width=0.9 \linewidth]{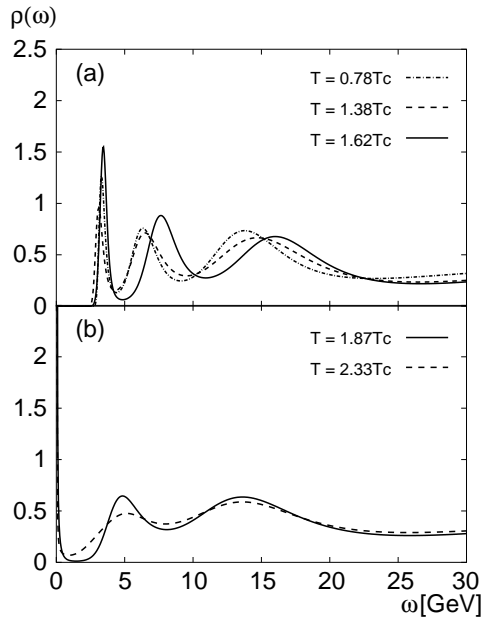}
\end{flushleft}
\vspace*{-0.8cm}
\caption{Spectral functions for $J/\psi$ (a) for
 $T/T_c=0.78, 1.38$, and $1.62$
 (b) for $T/T_c=1.87$ and $2.33$.}
\label{p-np}
\end{figure}

Shown in Fig.\ref{p-np}  are $\rho(\omega)$s \footnote{Following \cite{ahn01},
we define dimensionless
SPFs:  $A(\omega) = \omega^2 \rho(\omega)$ for $\eta_c$ and
$ A (\omega) = 3\omega^2 \rho(\omega)$ for $J/\psi$.}
for $J/\psi$ at $T/T_c = 0.78, 1.38$, and $1.62$ (Fig.\ref{p-np}(a))
and those at $T/T_c = 1.87$ and $2.33$ (Fig.\ref{p-np}(b))
(Corresponding $N_{\rm data}$ used in these figures
are $N_{\rm data}=89, ~40, ~34, ~33,$ and $25$
from low $T$ to high $T$).
If the deconfined plasma were composed of almost free
quarks and gluons, SPFs would show a smooth structure with
no pronounced peaks above the $q\bar{q}$ threshold.
To the contrary, we find a sharp peak near the zero temperature mass
even up to $T\simeq 1.6T_c$ as
shown in Fig.\ref{p-np}(a), while the peak 
disappears at $T\simeq 1.9 T_c$
as shown in Fig.\ref{p-np}(b).
The width of the first peak in Fig.\ref{p-np}(a) partly
reflects the unphysical broadening due to
the statistics of the lattice data and
partly reflects possible physical broadening
at finite $T$. At the moment, the former width of a few hundred MeV
seems to dominate and we are not able to draw
definite conclusions on the thermal mass shift and broadening.

\begin{figure}[hbt]
\begin{flushleft}
\includegraphics[width=0.8 \linewidth]{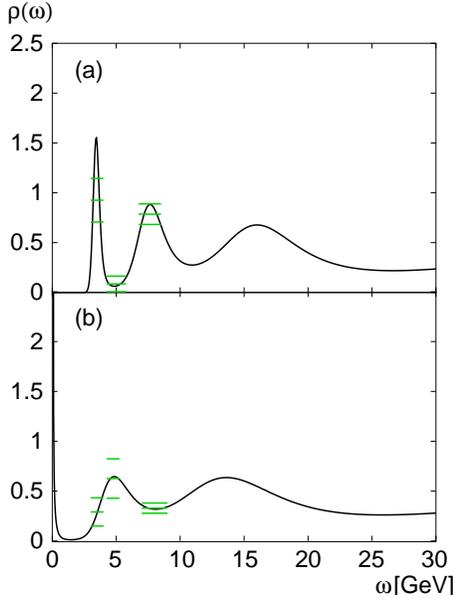}
\end{flushleft}
\vspace*{-0.8cm}
\caption{ Spectral functions for $J/\psi$ with MEM errors
(a) for $T/T_c=1.62$
(b) $T/T_c=1.87$.}
\label{error}
\end{figure}

Let us now evaluate the reliability of the existence (absence) of
the sharp peak at $T/T_c = 1.62 (1.87)$
by the two tests [I] and [II] mentioned before.
First test is the error analysis of the peak.
Shown in Fig.\ref{error} are the SPFs for $J/\psi$
at $T/T_c=1.62$ and 1.87 with MEM error bars
(The frequency interval over which
the SPF is averaged is characterized by the horizontal
position and extension of the bars, while the mean value and the 1$\sigma$
uncertainty of the integrated strength within the interval are
characterized by the heights of the bars).
The sharp peak at $T =1.62 T_c$ is statistically significant, and
the absence of the peak at the same position
at $T =1.87 T_c$ is also statistically significant.
The features found for $J/\psi$ in Figs. \ref{p-np} and \ref{error}
are also observed for $\eta_c$.

\begin{figure}[thb]
\begin{flushleft}
\includegraphics[width=0.9 \linewidth]{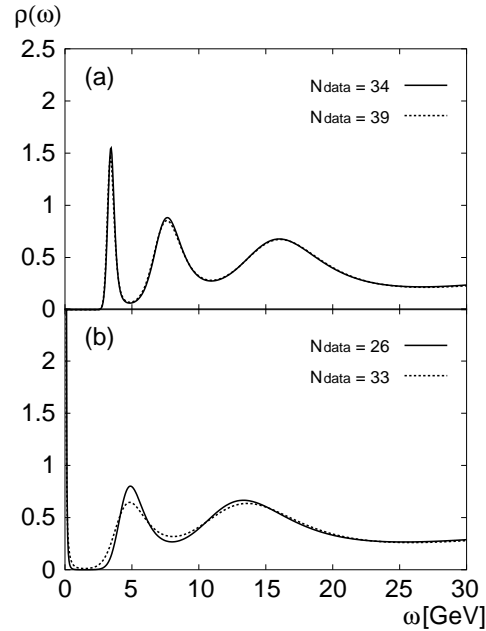}
\end{flushleft}
\vspace*{-0.8cm}
\caption{Comparison of the SPF for $J/\psi$
(a) for $ N_{\rm data} = 34$ and 39
  with $N_{\tau}=46$ $(T/T_c=1.62)$
(b) for $N_{\rm data} =26 $ and 33
  with $N_{\tau}=40$ $(T/T_c=1.87)$. }
\label{N-dep}
\end{figure}

The second test is the $N_{\rm data}$ dependence of the SPFs.
Shown in Fig.\ref{N-dep}(a) is the comparison of SPF
obtained with $N_{\rm data}=34$ and that of $N_{\rm data}= 39$
for the same temperature  $T =1.62 T_c$ $(N_{\tau} =46)$.
The two curves  are almost identical with each other.
Fig.\ref{N-dep}(b) shows SPFs obtained with
$N_{\rm data}=26$ and 33 for higher temperature
$T =1.87 T_c$ $(N_{\tau}=40)$.
Again the two curves are almost identical.
Therefore, the qualitative change of the SPF between
$T=1.62 T_c$ and 1.87$T_c$ is the real thermal effect  
and not caused by the artifact of the
insufficient number of data points.
     
To conclude, we have studied
the spectral functions of
$J/\psi$ and $\eta_c$ in the deconfined plasma
using lattice Monte Carlo data and the maximum entropy method.
The number of temporal sites $N_{\tau}$ is taken
as large as 46 and 40 for $T/T_c = 1.62$ and $1.87$, respectively.
Careful analyses of the MEM errors and the $N_{\rm data}$ dependence
of the results are carried out.
It is found that there are distinct resonances
up to $T \simeq 1.6T_c$ and they disappear between 1.6$T_c$
and $1.9 T_c$.

{\bf Acknowledgement} 
M.A. (T.H.) is partially
supported by the Grants-in-Aid of the Japanese Ministry of Education,
Science and Culture, No.~14540255 (No.~15540254).
Lattice calculations have been performed with the CP-PACS computer
under the ``Large-scale Numerical Simulation Program" of
Center for Computational Physics, University of Tsukuba.
 

\end{document}